\documentclass[prd,twocolumn,a4paper,superscriptaddress]{revtex4}
\usepackage{graphicx,amssymb}
\begin{document}
\newcommand{\be}{\begin{equation}}
\newcommand{\ee}{\end{equation}}
\newcommand{\bq}{\begin{eqnarray}}
\newcommand{\eq}{\end{eqnarray}}
\newcommand{\bsq}{\begin{subequations}}
\newcommand{\esq}{\end{subequations}}
\newcommand{\bc}{\begin{center}}
\newcommand{\ec}{\end{center}}
\newcommand {\R}{{\mathcal R}}
\newcommand{\al}{\alpha}
\newcommand\lsim{\mathrel{\rlap{\lower4pt\hbox{\hskip1pt$\sim$}}
    \raise1pt\hbox{$<$}}}
\newcommand\gsim{\mathrel{\rlap{\lower4pt\hbox{\hskip1pt$\sim$}}
    \raise1pt\hbox{$>$}}}

\title{Bouncing Eddington-inspired Born-Infeld cosmologies: an alternative to Inflation ?}

\author{P. P. Avelino}
\email[Electronic address: ]{ppavelin@fc.up.pt}
\affiliation{Centro de Astrof\'{\i}sica da Universidade do Porto, Rua das Estrelas, 4150-762 Porto, Portugal}
\affiliation{Departamento de F\'{\i}sica e Astronomia, Faculdade de Ci\^encias, Universidade do Porto, Rua do Campo Alegre 687, 4169-007 Porto, Portugal}
\author{R. Z. Ferreira}
\email[Electronic address: ]{ricardozambujal@gmail.com}
\affiliation{Departamento de F\'{\i}sica e Astronomia, Faculdade de Ci\^encias, Universidade do Porto, Rua do Campo Alegre 687, 4169-007 Porto, Portugal}

\date{\today}
\begin{abstract}

We study the dynamics of a homogeneous and isotropic Friedmann-Robertson-Walker universe in the context of the Eddington-inspired Born-Infeld theory of gravity. We generalize earlier results, obtained in the context of a radiation dominated universe, to account for the evolution of a universe permeated by a perfect fluid with an arbitrary equation of state parameter $w$. We show that a bounce may occur for $\kappa >0$, if $w$ is time-dependent, and we demonstrate that it is free from tensor singularities. We argue that Eddington-inspired Born-Infeld cosmologies may be a viable alternative to the inflationary paradigm as a solution to fundamental problems of the standard cosmological model.

\end{abstract}
\maketitle

\section{\label{intr}Introduction}

A new Eddington-inspired Born-Infeld (EiBI) theory of gravity \cite{Born:1934gh} has recently been proposed by Banados and Ferreira in \cite{Banados:2010ix}  (see also \cite{Deser:1998rj,Vollick:2003qp,Wohlfarth:2003ss,Nieto:2004qj,Comelli:2005tn,Vollick:2005gc,Zinoviev:2006,Ferraro:2008ey,Fiorini:2009ux,Ferraro:2009zk,Gullu:2010pc,Alishahiha:2010iq,Gullu:2010em,Casanellas:2011kf,DeFelice:2012hq,Delsate:2012ky} for other relevant studies of Born-Infeld type gravitational models and \cite{Clifton:2011jh} for a recent review on alternative theories of gravity). This theory is equivalent to Einstein's general relativity in vacuum but it leads to several attractive new features in the presence of matter. In particular, it has been shown that homogeneous and isotropic EiBI cosmologies may be singularity free and that non-singular compact stars may form even in the absence of pressure support (see \cite{Pani:2011mg,Avelino:2012ge,Pani:2012qb} for recent astrophysical and cosmological constraints).

The evolution of transverse, traceless (tensor) perturbations in the Eddington regime of a radiation dominated cosmology has been studied in \cite{EscamillaRivera:2012vz} and it has been shown that a linear instability develops as the universe approaches its maximum density. In \cite{Liu:2012rc} a 5-dimensional brane model has been considered as a possible solution to the stability of gravitational perturbations. In this paper we investigate the cosmological implications of the EiBI theory of gravity showing that there is a natural solution to the instability problem even in the absence of extra space-time dimensions.

In Sec. II we investigate the dynamics of a universe permeated by a perfect fluid with an arbitrary constant equation of state (EoS) parameter $w$, generalizing previous results obtained for $w=1/3$. We solve the equation for the evolution of tensor perturbations, showing that, independently of the value of $w$, there is an instability for both positive and negative $\kappa$. In Sec. III we consider a massive scalar field with a time-dependent EoS parameter and we show that a non-singular bounce may occur for $\kappa > 0$. We further argue that EiBI cosmologies provide an interesting alternative to inflation as a solution to some of the most fundamental problems of the standard cosmological model. We conclude in Sec. IV.

Throughout this paper we shall use fundamental units with $c=8\pi G=|\kappa|=1$ and a metric signature $(-,+,+,+)$. The Einstein summation convention will be used when a greek or latin index variable appears twice in a single term. Greek and latin indices take the values $0,...,3$ and $1,...,3$, respectively.

\section{EiBI cosmology (constant EoS parameter)}

The action for the Eddington inspired Born-Infeld theory of gravity is given by
\be
S=\frac{2}{\kappa}\int d^{4}x\left[\sqrt{\left|g_{\mu\nu}+\kappa R_{\mu\nu}\right|}-\lambda\sqrt{|g|}\right]+S_M\,,\label{eq:EddingtonBornInfeld Action}
\ee
where $g_{\mu\nu}$ are the components of the metric, $g$ is the determinant of $g_{\mu\nu}$, $R_{\mu\nu}$ is the symmetric Ricci tensor build from the connection  $\Gamma^\alpha_{\beta\gamma}$ and $S_M$ is the action associated with the matter fields. In the Palatini formulation $\Gamma^\alpha_{\beta\gamma}$ and  $g_{\mu\nu}$ are treated as independent fields. 

Varying the action with respect to the connection leads to the following equation of motion \
\begin{equation}
q_{\mu\nu}=g_{\mu\nu}+\kappa R_{\mu\nu}\,,\label{eq:ConnectionEquationOfMotion}
\end{equation}
where $q_{\mu\nu}$ is an auxiliary metric related to the original connection by $\Gamma^{\alpha}_{\beta\gamma} = {1 \over 2} q^{\alpha\zeta}(q_{\zeta\beta,\gamma} + q_{\zeta\gamma,\beta}- q_{\beta\gamma,\zeta})$ (a comma represents a partial derivative). Varying the action with respect to the metric one obtains the other equation of motion
\begin{equation}
\sqrt{|q|}q^{\mu\nu}=\lambda\sqrt{|g|}g^{\mu\nu}-\kappa\sqrt{|g|}T^{\mu\nu}\label{eq:MetricEquationOfMotion}
\end{equation}
where $q^{\mu\nu}$ is the inverse of $q_{\mu\nu}$. Without loss of generality we set $\lambda=1$ (this term can be included in the stress- energy tensor) and we shall use fundamental units with $|\kappa|=1$.

In order to analyse the cosmological implications of this theory we consider a flat homogeneous and isotropic Friedmann-Robertson-Walker universe with line element 
\be
ds^2=g_{\mu\nu}dx^\mu dx^\nu=-a^2 d\eta^2 +a^2\delta_{ij} dx^i dx^j \label{eq:FRWMetric}\,,
\ee
where $\eta$ is the conformal time, $x^i$ are comoving spatial coordinates (with $i=1,2,3$) and $\delta_{ij}$ is the Kronecker delta. We further assume that the universe is permeated with a perfect fluid with stress-energy tensor
\be
T^{\mu\nu}=\frac{1}{\sqrt {|g|}} \frac{\delta S_M}{\delta g_{\mu\nu}}=\left(p+\rho\right)u^{\mu}u^{\nu}+pg^{\mu\nu}\label{Tmunu}\,.
\ee
where $u^\mu$ are the components of the 4-velocity, $\rho$ is the energy density and $p$ is the pressure. The perfect fluid satisfies the standard conservation equation
\be
\dot{\rho}=-3\mathcal{H}\left(1+w\right)\rho\label{conserv}\,,
\ee
where $w=p/\rho$ is the EoS parameter (we shall assume that $w>-1$) and a dot represents a derivative with respect to $\eta$.

Writing $q_{\mu\nu}dx^\mu dx^\nu=-A^2 d\eta^2 +B^2\delta_{ij}dx^i dx^j$ one obtains, using Eq. (\ref{eq:MetricEquationOfMotion}), that
\bq
A^2&=&a^2 \frac{D}{1+{\kappa \rho}}\,,\\
B^2&=&a^2  \frac{D}{1-w {\kappa \rho}}\,,
\eq
where $D=(1+{\kappa \rho})^{1/2}(1-w{\kappa \rho})^{3/2}$.

The Friedmann equation can be obtained from Eqs. (\ref{eq:ConnectionEquationOfMotion}) and (\ref{eq:MetricEquationOfMotion}). It is given by
\be
\mathcal{H}^{2}=\frac{8}{3}a^2\frac{F_1}{F_2^2}\label{eq:Friedmann Equation}\,,
\ee
where the functions $F_1$ and $F_2$ are equal to
\bq
F_1&=&\kappa  \left(1+{\kappa \rho}\right)\left(1-{\kappa \rho} w\right)^2 \times \nonumber \\
&\times&\left(-2+{\kappa \rho}(1+3w)+2D\right)\,,\\
F_2&=&4+{\kappa \rho}\left(1-2w\left(2-{\kappa \rho}\right)+3w^2\left(1+2{\kappa \rho}\right)\right)\,.
\eq
For $w=1/3$ one recovers the result found in \cite{Banados:2010ix}.

At low densities one obtains
\be
\mathcal{H}^{2}=\frac{\rho a^2}{3}+\frac{a^2 \kappa \rho^2}{8}\left(1+w\right) \left(1-3w\right)\label{exp0}\,,
\ee
up to second order in $\kappa \rho$. 

According to Eq. (\ref{eq:Friedmann Equation}), $\mathcal{H}=0$ when $\kappa=-1$ and $\rho=1$. Near this critical point one has
\be
\mathcal{H}^{2}=\frac{8w^{2}a^{2}}{9\left(1+w\right)}\left(1-\rho\right)+O\left((1-\rho)^{3/2}\right)\label{expm1}\,.
\ee
If $\kappa=-1$  then Eqs. (\ref{conserv}) and (\ref{expm1}) imply that a bounce occurs when $\rho=1$, with $1-\rho \propto \eta^2$ for $\rho \sim 1$ (taking $\eta=0$ at the bounce). Hence, in the absence of perturbations, EiBI cosmologies are singularity free for $\kappa<0$ and the universe has no Big Bang.

On the other hand, if ${\dot w}=0$, $\kappa=1$ and $w>0$ then there is another critical point with $\mathcal{H}=0$ at $\rho=w^{-1}$. Expanding around this point, one obtains
\be
\mathcal{H}^{2}=\frac{8a^{2}w^{2}}{27\left(1+w\right)^{2}}\left(\frac{1}{w}-\rho\right)^{2}+O\left(\left(\frac{1}{w}-\rho\right)^{3}\right)\label{expm1ow}\,.
\ee
If  $\kappa=1$ then Eqs. (\ref{conserv}) and (\ref{expm1ow}) imply that $\rho \to w^{-1}$ when $\eta \to -\infty$, with $1-w\rho \propto \exp\left(\eta\right)$ \,.
Consequently, if $\kappa > 0$ then the singularity may be avoided in the case of a constant EoS parameter $w>0$, with the universe becoming static when $\eta \to - \infty$.

\subsection{Tensor modes}

Consider a perturbed homogeneous and isotropic flat space time characterized by 
\bq
g_{\mu\nu}dx^\mu dx^\nu&=&-a^2 d\eta^2 +a^2\left(\delta_{ij} +h_{ij}\right) dx^i dx^j \label{gmunu}\,, \\
q_{\mu\nu}dx^\mu dx^\nu&=&-A^2 d\eta^2 +B^2 \left(\delta_{ij}+\gamma_{ij}\right)dx^i dx^j \label{qmunu}\,,
\end{eqnarray}
where $a$, $A$ and $B$ are functions of the conformal time $\eta$ and $h_{ij}$ is transverse and traceless ($h_{ii}=0$ and ${h^{ij}}_{,i}=0$). In \cite{EscamillaRivera:2012vz} it has been shown that $\gamma_{ij}=h_{ij}$ and that the dynamics of $h_{ij}$ is given by
\be
{\ddot h}_{ij}+F_4 {\dot h}_{ij}+F_5 k^2h_{ij}=0\,. \label{hijdd}
\ee
where
\bq
F_4&=&3\frac{\dot B}{B}-\frac{\dot A}{A}=2\mathcal{H}+\frac{\kappa{\dot \rho}}{1+{\kappa \rho}}\,, \label{F4}\\
F_5&=&\left(\frac{A}{B}\right)^2=\frac{1-w{\kappa \rho}}{1+{\kappa \rho}}\label{F5}\,,
\eq
and $k$ is a comoving wavenumber.

\subsubsection{$\kappa=-1$}

If $\kappa = -1$ then the universe has a bounce for $\rho=1$. Near the bounce, $\mathcal{H} \to 0$ and the infinite wavelength ($k=0$) limit of Eq. (\ref{hijdd}) can be written as
\be
\frac{d{\dot h}_{ij}}{d\rho}-\frac{{\dot h}_{ij}}{1-\rho}=0\,, \label{hijdd1}
\ee
which implies that ${\dot h}_{ij} \propto (1-\rho)^{-1}$. Dividing Eq. (\ref{hijdd1}) by $\dot \rho$ one finds, using Eqs.  (\ref{expm1}) and (\ref{F4}), that
\be
h_{ij} \propto (1-\rho)^{-1/2}+C_1\,,
\ee
where $C_1$ is a real constant. This signals the presence of a singularity at the bounce for any value of $w$ if $\kappa=-1$. We shall see that this singularity cannot be avoided by considering ${\dot w} \neq 0$, as long as $w$ is well behaved at the bounce.

\subsubsection{$\kappa=1$}

If $\kappa=1$ then, for $w > 0$, $\rho \to w^{-1}$ when $\eta \to -\infty$ (for $w \le 0$ the singularity is not avoided even in the absence of cosmological perturbations). Hence, both $F_4$ and $F_5$ tend to zero as $\eta \to -\infty$. In that limit Eq. ({\ref{hijdd}) reduces to ${\ddot h}_{ij}=0$, whose solution is $h_{ij} \propto C_2 \eta +C_3$ ($C_2$ and $C_3$ are real constants), signaling the presence of an instability in the infinite past \cite{EscamillaRivera:2012vz}. 

\section{EiBI cosmology (variable EoS parameter)}

In the previous section we generalized, for an arbitrary constant value of $w>-1$, the results presented in \cite{Banados:2010ix,EscamillaRivera:2012vz} which were obtained considering $w=1/3$. In this section we further generalize our analysis to include models where ${\dot w} \neq 0$. A time dependent $w$ is expected, for example, if the energy density of the universe is dominated by a dynamical scalar field, in particular in the transition between different cosmological epochs \cite{Avelino:2009ze}. If ${\dot w} \neq 0$ then Eq. (\ref{eq:Friedmann Equation}) is modified to
\be
\mathcal{H}^{2}=\left(\frac{a{\sqrt F_1}+F_3\dot{w}}{F_2}\right)^{2} \label{eq:Friedmann Equationvw}\,,
\ee
where
\be
F_3=-3\rho\left(1+{\kappa \rho}\right)\,.
\ee
If $\kappa=-1$ and $\rho=1$ then both $F_1$ and $F_3$ vanish, which implies that $\mathcal{H}=0$ for $\rho=1$, even for ${\dot w} \neq 0$.  On the other hand, if $\kappa=-1$ then the term $F_3 {\dot w}$ becomes negligible compared to $a {\sqrt F_1}$ in the $\rho \to 1$ limit, assuming that ${\dot w}$ is finite near the bounce at $\rho=1$. Consequently, the tensor instability, previously found for a constant $w$ assuming $\kappa=-1$, cannot be avoided even for a time-dependent $w$. In the remainder of this paper we shall focus on the more interesting $\kappa=1$ case. For the sake of completeness it is important to mention that if ${\dot w} \neq 0$ then Eq. (\ref{exp0}) should be modified to
\bq
\mathcal{H}^{2}&=&\frac{\rho a^2}{3}-\frac{a\dot{w} \kappa \rho^{3/2}}{2\sqrt{3}} +\nonumber\\
&+&\frac{a^2 \kappa \rho^2}{16}\left(2\left(1+w\right) \left(1-3w\right)+\frac{\kappa\dot{w}^{2}}{a^{2}}\right)\label{exp0vw}\,,
\eq
while Eq. (\ref{expm1}) remains valid.

\subsection{Scalar field}

Let us assume that the matter fields can be described by a real scalar field with action
\be
\label{eq:L}
S_M=\int d^4x \, \sqrt{|g|} \mathcal \, {\cal L} \, ,
\ee
where
\bq
{\cal L}&=&{\cal L}(\phi,X)\,, \\
X&=&-\frac{1}{2}\phi^{,\mu}  \phi_{,\mu} \label{eq:kinetic_scalar1}\,.
\eq
For time-like $\phi_{,\mu}$, the energy-momentum tensor associated with a scalar field $\phi$ may be written in a perfect fluid form by means of the following identifications
\be
u_\mu = \frac{\phi_{,\mu}}{\sqrt{2X}} \,,  \quad \rho = 2 X {\mathcal L}_{,X} - {\mathcal L} \, ,\quad p =  {\mathcal L} \label{eq:new_identifications}\,,
\end{equation}
so that the EoS parameter is given by
\be
w=\frac{\mathcal L}{2X{\mathcal L}_{,X}-{\mathcal L}}\,.
\ee

In this paper we shall consider a standard scalar field with lagrangian given by
\be
{\mathcal L}=X-V(\phi)\,,
\ee
where $V$ is the scalar field potential. The value of $w$ is given by
\be
w=\frac{X-V}{X+V}\,,
\ee
where $X=a^2 {\dot \phi}^2/2$. Here, we shall consider a massive scalar field with $V(\phi)=m^2\phi^2/2$ so that $w$ may vary continuously between $-1$ and $1$.

\begin{figure}
\includegraphics*[width=8.0cm, height=8.0cm]{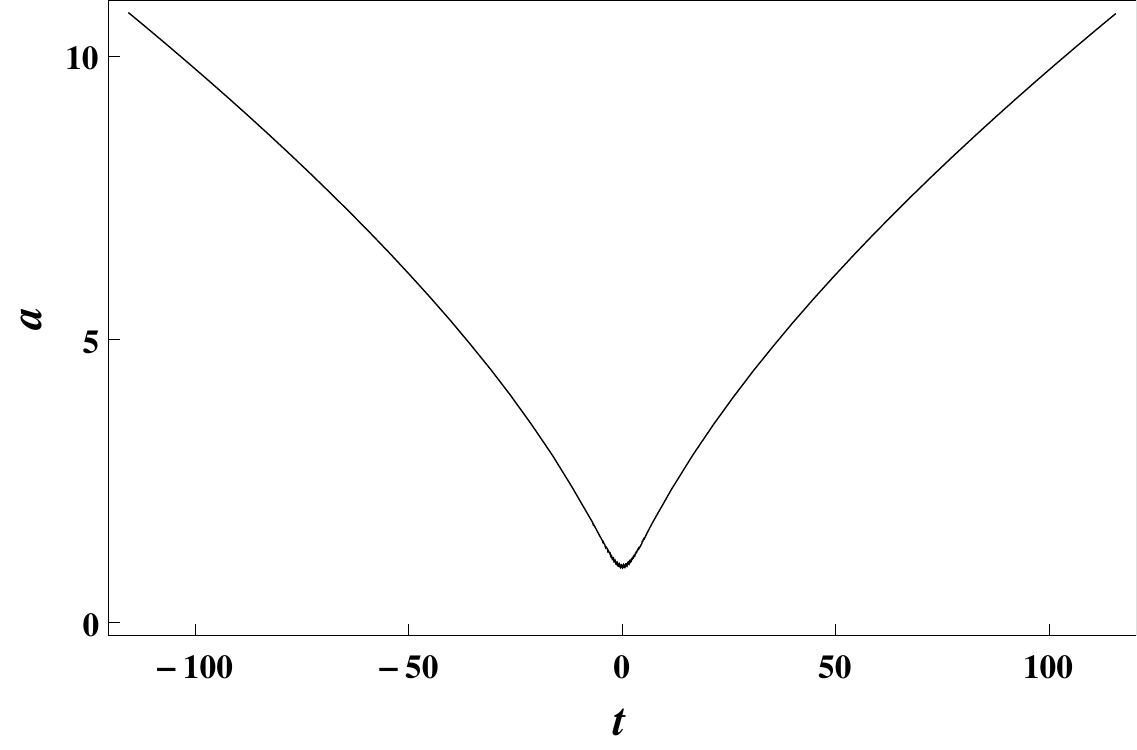}
\caption{\label{Fig1} Evolution of $a$ as a function of the physical time $t$ near the bounce.}
\end{figure}

\subsection{Bounce with $\kappa=1$}

If $\kappa=1$ and ${\dot w}\neq 0$ then the condition $\mathcal{H}=0$ is no longer verified for $\rho=w^{-1}$. Instead, depending on the value of ${\dot w}$, this critical point might occur for different values of $w$. Unlike the constant $w > 0$ case, where $\rho \to w^{-1}$ as $\eta \to -\infty$ (with $\dot {\mathcal{H}} \to 0$ as $\mathcal{H} \to 0$), if ${\dot w} \neq 0$, in general, $\dot {\mathcal{H}} \neq 0$ when $\mathcal{H} = 0$. This implies that a bounce may occur. Here, we present a simple example with a real scalar field considering initial conditions with $\rho_i=10^{-4}$ and $w_i=0$. Fig. 1 shows the evolution of $a$ as a function of the physical time  $t=\int_{1}^a {\mathcal{H}(a')}^{-1} da'$ near the bounce for $m=100$ (the scale factor is normalized to unity at the bounce). The presence of a bounce in the $\kappa=1$ case removes the tensor instability since $F_4 \sim 0$ near the bounce while for $\rho \ll 1$ the dynamics of $h_{ij}$ becomes identical to the one obtained in Einstein's General Relativity (with $F_4=2\mathcal{H}$ and $F_5=1$).

\subsection{An alternative to Inflation}

Bouncing EiBI cosmologies may be an interesting alternative to inflation (see \cite{Linde:2007fr,Mazumdar:2010sa} for recent reviews of cosmic inflation) as a solution to fundamental problems of the standard cosmological model. The singularity problem may be solved in this class of models since there is no big bang singularity if $\kappa=1$. If the contracting period is sufficiently long then the age of the universe as well as its particle horizon can be arbitrarily large and consequently there is no horizon problem. Given that the universe can be arbitrarily large at early times, the size age and entropy problems do not arise. The curvature problem is significantly alleviated since the universe approaches the critical density in the contracting phase. The monopole problem can also be solved if the bounce occurs at an energy scale considerably smaller than the Grand Unification Scale.  In that case the Planck era is also avoided. In addition, a nearly scale invariant spectrum of density fluctuations may also be realized in the context of the ``matter bounce'' scenario, leading to a scale-invariant spectrum of perturbations generated in a matter dominated collapsing phase \cite{Wands:1998yp,Finelli:2001sr} (see also \cite{Novello:2008ra,Brandenberger:2012uj} for a detailed discussion of bouncing cosmologies). 

\section{\label{conc}Conclusions}

In this paper we generalized earlier results for the dynamics of EiBI cosmologies to accommodate an arbitrary constant EoS parameter $w$. In the absence of fluctuations, we have shown that the singularity problem can be avoided if $\kappa < 0$ or if $\kappa > 0$ and $w>0$. However, we have demonstrated that the inclusion of linear tensor perturbations leads to an instability which cannot be lifted for $\kappa<0$, even if ${\dot w} \neq 0$. Still, we have shown that a time-dependent $w$ can naturally lead to a bounce which is free from cosmological singularities if $\kappa > 0$. We have argued that several of the fundamental problems of the standard cosmological model can be solved in the context of Eddington-inspired Born-Infeld gravity even in the absence of an inflationary era in the early universe.



\bibliography{EiBI}

\end{document}